\documentclass[manuscript]{aastex}
\usepackage{CJK}

\usepackage{graphicx}

\shorttitle{Disentangling the Entangled: VLA1623}
\shortauthors{Murillo \& Lai}

\begin{document}
\begin{CJK*}{UTF8}{bsmi}

\title{Disentangling the entangled: Observations and analysis of the triple non-coeval protostellar system VLA1623}

\author{Nadia M. Murillo (穆美蓉)\altaffilmark{1} and Shih-Ping Lai (賴詩萍)\altaffilmark{1,2}}

\altaffiltext{1}{Institute of Astronomy and Department of Physics, National Tsing Hua University, 101 Section 2 Kuang Fu Road, Hsinchu 30013, Taiwan}
\altaffiltext{2}{Academia Sinica Institute of Astronomy and Astrophysics, P. B. Box 23-141, Taipei 10617, Taiwan}

\begin{abstract}
Commonplace at every evolutionary stage, Multiple Protostellar Systems (MPSs) are thought to be formed through fragmentation, but it is unclear when and how. The youngest MPSs, which have not yet undergone much evolution, provide important constraints to this question. It is then of interest to disentangle early stage MPSs. In this letter we present the results of our work on VLA1623 using our observations and archival data from the Submillimeter Array (SMA). Our continuum and line observations trace VLA1623's components, outflow and envelope, revealing unexpected characteristics. We construct the SED for each component using the results of our work and data from literature, as well as derive physical parameters from continuum and perform a simple kinematical analysis of the circumstellar material. Our results show VLA1623 to be a triple non-coeval system composed of VLA1623A, B \& W, with each source driving its own outflow and unevenly distributed circumstellar material. From the SED, physical parameters and IR emission we conclude that VLA1623A \& W are Class 0 and I protostars, respectively, and together drive the bulk of the observed outflow. Furthermore, we find two surprising results, first the presence of a rotating disk-like structure about VLA1623A with indications of pure Keplerian rotation, which, if real, would make it one of the first evidence of Keplerian disk structures around Class 0 protostars. Second, we find VLA1623B to be a bonafide extremely young protostellar object between the starless core and Class 0 stages.
\end{abstract}


\maketitle
\end{CJK*}

\section{Introduction}
Multiple protostellar systems (MPSs) are common among every stage of protostellar evolution, but are tricky to decipher. Fragmentation is believed to be the main mechanism for forming MPSs \citep{toh02}, but how and when fragmentation occurs is still a major question. To constrain the fragmentation process, information about the system's multiplicity, coevality, environment and outflows are needed. Since older systems have suffered dynamical evolution \citep{goo07}, young early stage MPSs, having many of the initial conditions almost intact, are the best targets for the task. In this letter we present the results of our work on the much debated VLA1623, located in $\rho$ Ophiuchus and generally considered the prototypical Class 0 source (\citealt{and93}, hereafter AWB93).

First detected through red- and blueshifted overlapped outflows \citep{and90}, VLA1623's multiplicity has long been suspected from observation of multiple continuum sources (\citealt{leo91}; \citealt{str95}; \citealt{lmw00}, hereafter LMW00), outflow misalingment (\citealt{den95}; \citealt{yu97}, hereafter YC97) and two discreet precessing jets originating from the same region \citep{den95, cog06}. In contrast, the multiple continuum sources are argued to constitute a monopolar jet driven by VLA1623A (\citealt{bon97}, hereafter BA97; \citealt{mau10}; \citealt{mau12}, hereafter MOA12). While VLA1623A is agreed to be a real protostellar source, and was used by ABW93 as the prime example to establish a new class of YSOs - ``Class 0", there is still debate on the nature of the other sources, mainly VLA1623B (\citealt{wkga11}, hereafter WKGA11).

We present the results of our work to determine the multiplicity of VLA1623 and disentangle its outflows. Our own Submillimeter Array (SMA; \citealt{ho04}) observations and archival data were used, suplemented with data from literature. Our results show VLA1623 to be a triple non-coeval protostellar system and find VLA1623B to be a bona-fide protostar with an extremely young age.

\section{Data}
\subsection{Observations}
We observed VLA1623 with SMA at 345GHz in the compact configuration for 8 hours on 2007 July 25, with 2GHz correlator mode configured to observe $^{12}$CO (3-2), C$^{17}$O (3-2) and SO (9-8) simultaneously with 870$\mu$m continuum and polarization measurements. All windows were set to have 128 channels, providing a velocity resolution of 0.704\,km\,s$^{-1}$. Quasars 3c273 and 1733-130 were observed as the bandpass and gain calibrators. A planet was not observed for unknown operational reasons. This track's original purpose was polarization measurements, where lack of flux calibration is not a critical issue. We performed standard data reduction for polarization data using MIRIAD. No significant polarization detection was made. Flux calibration is done with 1733-130 scaled to 1.31Jy, value taken from the SMA Observer Center, with measured uncertainty of 0.035Jy. Imaging with natural weighting resulted in a beamsize of 2\farcs36$\times$1\farcs41. 

We further observed VLA1623 at 345GHz in the extended configuration with a 3-hr filler track on 2012 February 14 using 4GHz correlator mode. Only the continuum emission is considered. Calibration was performed on IDL MIR package following the standard calibration pipeline. 3c84, 1733-130 and Callisto were used for bandpass, gain and flux calibration. Calibrated flux of 1733-130 is 1.0$\pm$0.006Jy, in agreement with the flux listed on SMA Observer Center. Data imaging and analysis were performed with MIRIAD using natural weighting resulting in a beamsize of 1\farcs51$\times$0\farcs76.

We compared the two datasets by binning vectorially averaged amplitudes, and when the tracks overlap, the flux level was comparable. Uncertainty in flux calibration is typically $\sim$10\%.

\subsection{Archival Data}
VLA1623 was observed at 230GHz in the very extended configuration on 2007 June 11 and 2009 July 3. The 2009 data were presented by MOA12. Here we redo the data reduction and analysis. The data quality of the 2007 data appears to be better than the 2009 data; the system temperatures are 92$\pm$24K and 130$\pm$39K, and the atmospheric opacities are 0.06 and 0.1 for the 2007 and 2009 data, respectively. Both observations used the 2GHz correlator mode configured for $^{12}$CO (2-1), C$^{18}$O (2-1), $^{13}$CO (2-1) and continuum. The three windows with line emission in both datasets were set to have 512 channels each, giving a velocity resolution of 0.3\,km\,s$^{-1}$. The remaining windows were configured to have 32 channels each. 

For the 2007 data, the quasar 3c454.3 for bandpass, 1514-241 and 1622-297 for gain, and Callisto for flux were used in calibration. For the 2009 data, bandpass and gain calibration was done with 1924-292 and 1625-254. Flux calibration using Callisto caused the flux of VLA1623B to rise 30mJy compared to the 2007 track. We suspect a glitch occured during Callisto's observation, so 1625-254 was used as a flux calibrator instead with 0.79Jy taken from the SMA Observer Center around the observing date, with measured uncertainty of 0.025Jy. After scaling, the flux from the two datasets are consistent within 11\%. MIRIAD was used to separately calibrate each dataset and then combine during imaging assuming natural weighting in order to increase the Signal-to-Noise ratio in continuum and $^{12}$CO. The resulting beamsize is of 0\farcs61$\times$0\farcs43. Only in the 2007 data C$^{18}$O showed detection above 3$\sigma$ and neither dataset detected $^{13}$CO.

\section{Results}
\subsection{Continuum}
\begin{figure}
\centering
\includegraphics[angle=-90,scale=0.7]{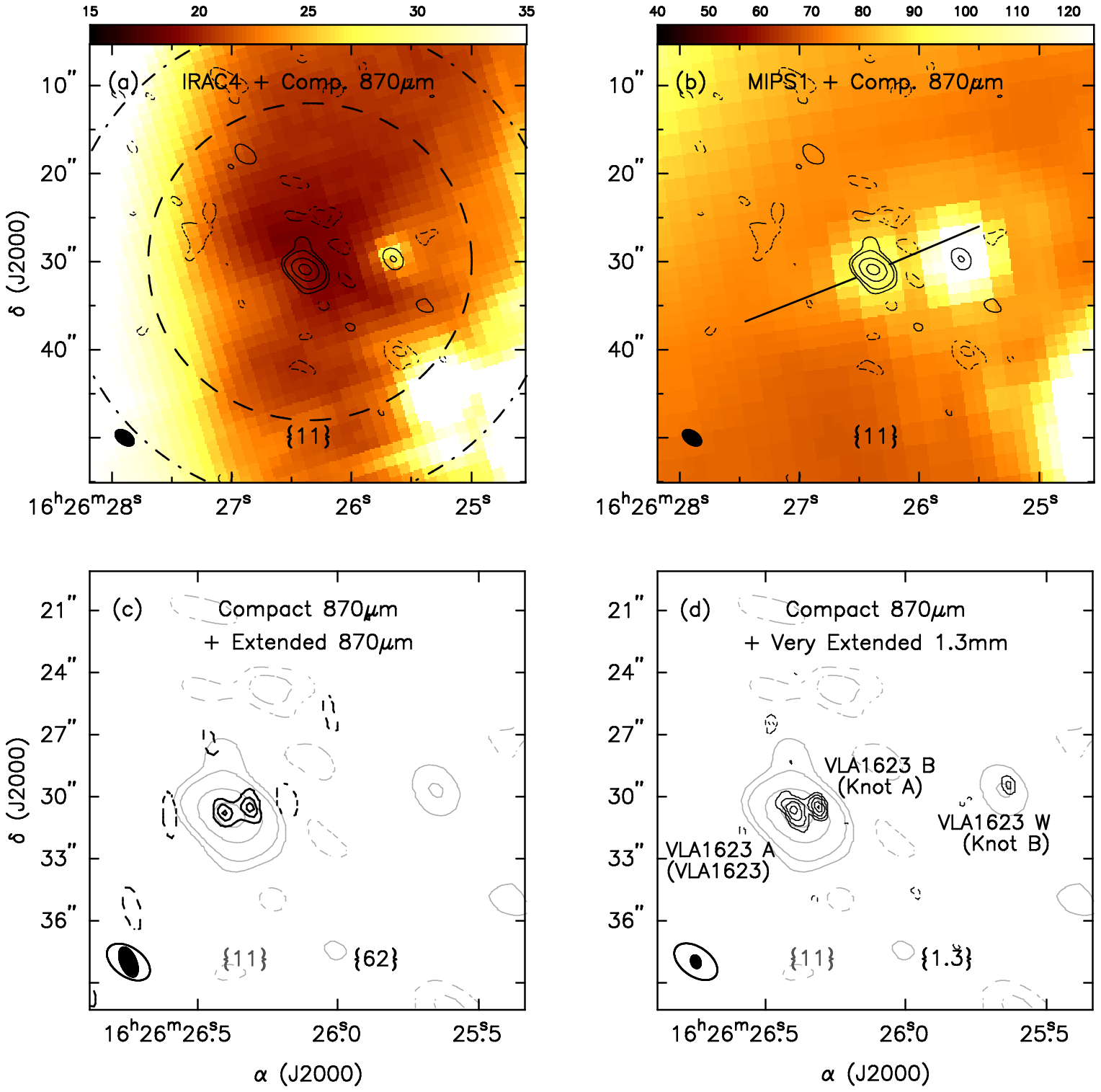}
\caption{Continuum: Spitzer (a) IRAC4 (8$\mu$m) and (b) MIPS1 (24$\mu$m) halftone images; (c) Extended 870$\mu$m continuum; (d) Very Extended 1.3mm continuum with alternate component names. All images are overlaid with 870$\mu$m compact continuum contours (light grey in (c) and (d)) for scale comparison. Dashed and dashed-dot circles show the SMA primary beam for 870$\mu$m and 1.3mm, respectively. RMS noise (mJy\,beam$^{-1}$) is in curly brackets. Contours are in steps of -5$\sigma$, -3$\sigma$, 3$\sigma$, 5$\sigma$, 10$\sigma$, 30$\sigma$ and 60$\sigma$ for 870$\mu$m compact continuum; -4$\sigma$, -3$\sigma$, 3$\sigma$, 4$\sigma$ and 5$\sigma$ for 870$\mu$m extended continuum; and -5$\sigma$, -3$\sigma$, 3$\sigma$, 5$\sigma$, 10$\sigma$, 20$\sigma$ and 50$\sigma$ for 1.3mm Very extended continuum. (Color figure in the online journal)}
\label{figcont}
\end{figure}

Compact 870$\mu$m data shows two continuum peaks (Figure~\ref{figcont}). The unresolved stronger peak coincides with LMW00's VLA1623A \& B. A weaker peak located 10\farcs5 to the west, matches with BA97's knot B. We will refer to this west peak with YC97's notation VLA1623W. Hereafter we refer to VLA1623A, VLA1623B and VLA1623W as A, B and W, respectively.

Extended 870$\mu$m observations resolve A \& B, but W is too weak to be detected in such short observations. B appears brighter than A possibly due to B being more compact than A and that much of the emission from A, present in the compact data, has been resolved out (Figure~\ref{figcont}c). 

The extended 1.3mm continuum observations resolved three peaks coinciding with the positions of A, B and W. A presents an elongated flattened structure perpendicular to the outflow axis, while B is compact and W is much dimmer (Figure~\ref{figcont}d). 

\begin{figure}
\centering
\includegraphics[scale=0.5]{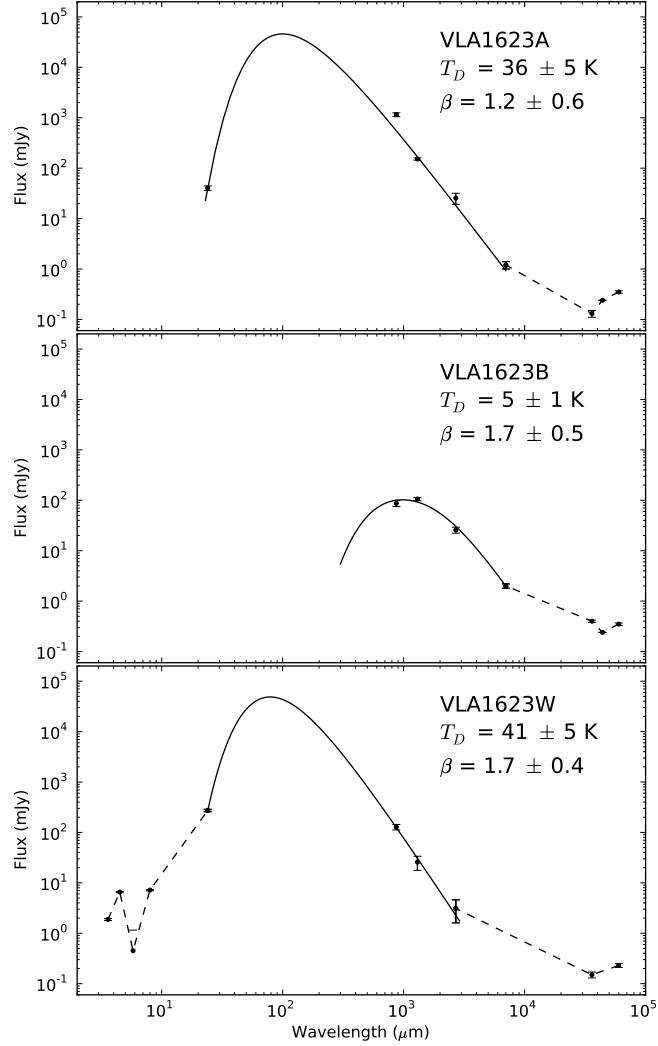}
\caption{SED for each component of VLA1623 constructed using data from BA97, LMW00, WKGA11 and this work (Table~\ref{tball}). The solid black line shows the best single-temperature graybody fit for fluxes at 7mm$\ge$$\lambda$$\ge$24$\mu$m, with the best fit dust temperature and dust emissivity index $\beta$ written in the upper right corner. A simple interpolation (dashed line) is made for points outside the fitting range.}
\label{figsed}
\end{figure}

We identify the infrared counterparts by comparing 870$\mu$m and 1.3mm continuum to Spitzer data (Figure~\ref{figcont}a,b). Unresolved A \& B are only detected in the MIPS1 band (40.4$\pm$4.24mJy) with the emission centered at A. Since B's SED starts to drop at 870$\mu$m (Figure~\ref{figsed}) we assume the MIPS1 emission is from A. Lack of detection of A \& B in all IRAC bands indicates that both components are still deeply embedded and very young. Surprisingly we find that W coincides with a bright infrared source detected in the four IRAC bands and MIPS1 band. Resolved fluxes and uncertainties for each component are listed in Table~\ref{tball}.

We construct each component's SED using all available resolved fluxes from BA97, LMW00, WKGA11 and our work (Table~\ref{tball}). The source size in all literature are in the 1$\sim$2$\arcsec$ range except in MIPS1 band; therefore, the flux we used originates from similar spatial scales. The MIPS1 detections are point sources, so their large source size is due to poor resolution. Fluxes between 24$\mu$m$\le$$\lambda$$\le$7mm were fit by a single-temperature greybody fit with optical depth $\tau$$\propto$$\nu$$^{\beta}$. The best fit curves are shown in Figure~\ref{figsed} and fitting results are listed in Table~\ref{tball}. The uncertainty in the absolute flux calibration may affect our fitting results. If we adopt the most conservative estimate for SMA flux uncertainty ($\sim$30\%, \citealt{ho04}), the derived parameters (Table~\ref{tball}) will only be altered by $\lesssim$10\%.

Using the dust temperature, T$_{D}$, obtained from the SED fitting, assuming $\kappa$$_{230}$\,=\,0.01\,cm$^{2}$\,g$^{-1}$ and a distance of 120$\pm$4.5\,pc \citep{loi08}, we derive the total (gas and dust) circumstellar mass M$_{cir}$, column and number density (N$_{H_{2}}$ \& n$_{H_{2}}$) for each component from 1.3mm continuum (Table~\ref{tball}, uncertainties were obtained through error propagation accounting for errors in distance, flux and T$_{D}$). Using the fluxes listed in Table~\ref{tball} we derive the bolometric and submillimeter luminositites (L$_{bol}$ \& L$_{submm}$) and their ratio for each component. We find L$_{submm}$/L$_{bol}$\,=\,1.17\% and 0.18\% for A and W, ratios characteristic of Class 0 and Class I sources, respectively \citep{you05, fro05}. For B we find L$_{bol}$\,=\,8.9$\times$10$^{-5}$L$_{\odot}$, considerably lower than VeLLOs.

\cite{dun08} showed that incomplete SEDs lead to underestimated derived luminosities. To examine whether this affects A \& W's classification,  we calculate the luminosities from the best fit SED curve including the area under the extrapolated curve, L$_{SED}$, and then compare with the above derived luminosities (Table~\ref{tball}). We find that L$_{SED}$ are higher than the derived luminosities, as predicted, however the ratios are consistent with the evolutionary classification previously obtained.

\subsection{Outflow}
\begin{figure}
\centering
\includegraphics[angle=-90,scale=0.6]{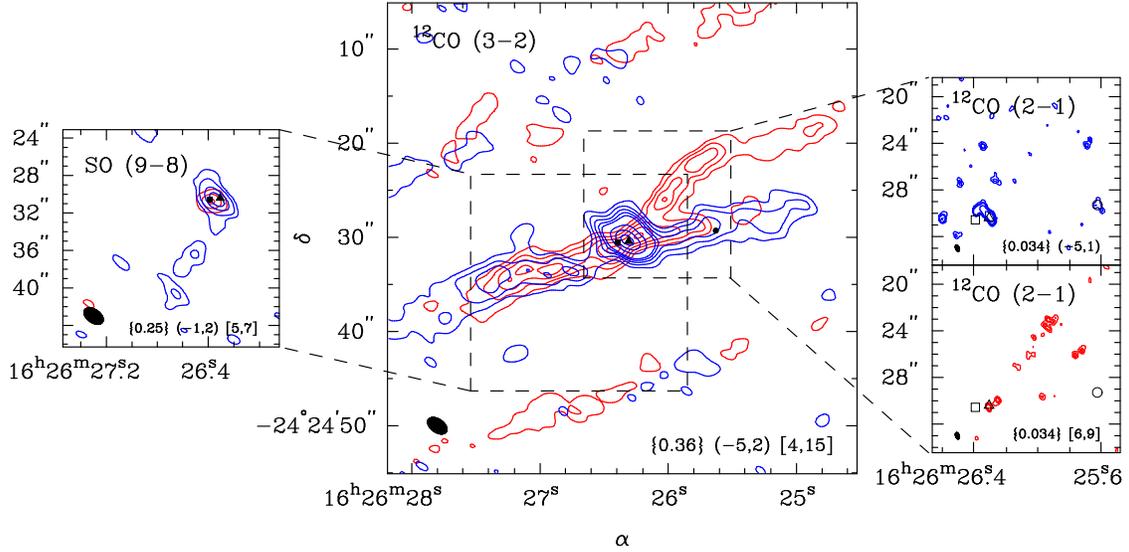}
\caption{Intensity integrated maps for SO (9-8), $^{12}$CO (3-2) and $^{12}$CO (2-1). Redshifted and blueshifted emission are shown in black and grey contours, respectively. RMS noise (Jy\,beam$^{-1}$\,km\,s$^{-1}$), integrated blue- and redshifted velocity ranges (km\,s$^{-1}$) are listed in curly, round and square brakets, respectively.  Contours are in steps of 10$\sigma$, 20$\sigma$, 30$\sigma$, 40$\sigma$, 50$\sigma$, 90$\sigma$, 130$\sigma$ and 170$\sigma$ for $^{12}$CO (3-2); 4$\sigma$, 6$\sigma$, 7$\sigma$, 12$\sigma$ and 25$\sigma$ for $^{12}$CO (2-1); and 3$\sigma$, 6$\sigma$, 12$\sigma$ and 18$\sigma$ for SO (9-8). Negative contours not plotted for clarity. Positions of VLA1623A, B \& W are marked with a square, triangle and circle, respectively. (Color figure in the online journal)}
\label{figout}
\end{figure}

In Figure~\ref{figout}, we show  VLA1623's outflows in $^{12}$CO (3-2) and (2-1) with a velocity range of -6--2\,km\,s$^{-1}$ and 4--15\,km\,s$^{-1}$ for the blueshifted and redshifted emission, respectively (v$_{lsr}$ =3.8kms$^{-1}$, YC97). Overlapped red- and blueshifted morphologies are observed in both transitions, as expected (\citealt{and90}; \citealt{den95}; YC97), however we observe several additional features not previously reported. In $^{12}$CO (3-2) the SE lobe is collimated with overlapped emission throughout, whereas the NW lobe presents a forked structure with overlapped red- and blueshifted emission only in the southernmost branch. In $^{12}$CO (2-1) a slow blueshifted feature is located at W while a dim redshifted clumpy ouflow coincides with the northernmost branch of the NW lobe. In both transitions we observe a compact blueshifted knot elongated in the NE to SW direction and centered on B. The blueshifted knot presents a spatially overlapped redshifted counterpart only in $^{12}$CO (2-1) (Figure~\ref{figout}).

SO spatially overlapped blueshifted (-0.5--2.3\,km\,s$^{-1}$) and redshifted (5--7.2\,km\,s$^{-1}$) emission as well as a systemic velocity component (3--4.4\,km\,s$^{-1}$) and a southward blueshifted ridge are detected around A \& B (Figure~\ref{figout}). The SO emission is probably tracing circumstellar material shocked by A \& B's outflows.

\subsection{Circumstellar Envelope}
\begin{figure}
\centering
\includegraphics[angle=-90,scale=0.7]{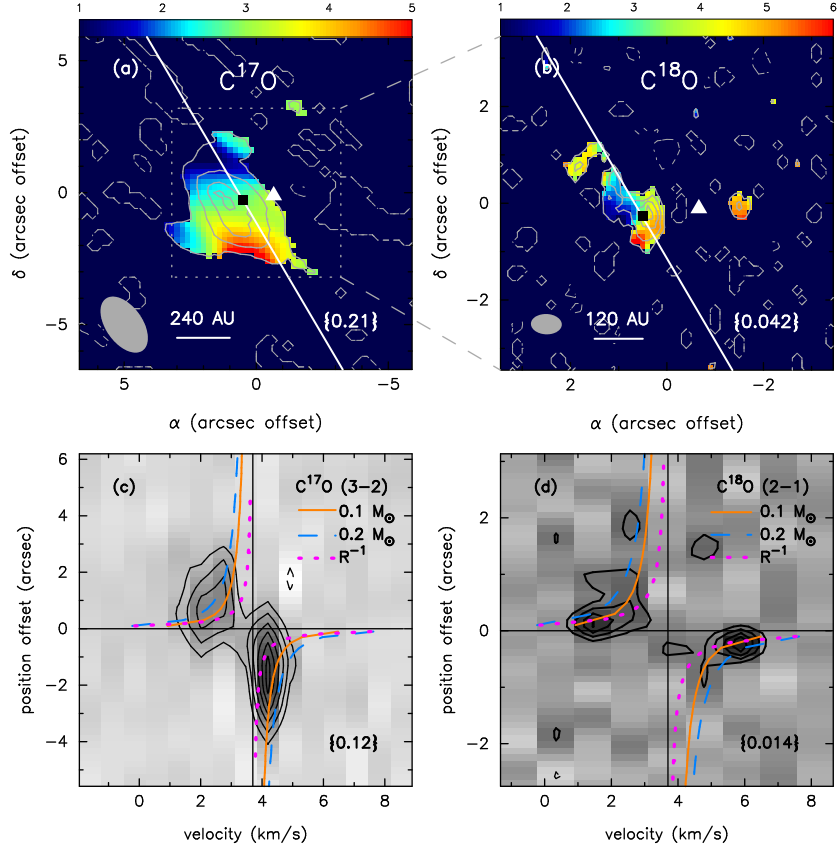}
\caption{Intensity integrated (contours) and Velocity gradient (halftone) maps for (a) C$^{17}$O and (b) C$^{18}$O. The diagonal line in both maps indicates the Position-Velocity (PV) diagram cut (P.A.\,=\,30$^\circ$) for (c) C$^{17}$O and (d) C$^{18}$O. RMS noise (Intensity integrated: Jy\,beam$^{-1}$\,km\,s$^{-1}$; PV diagram: Jy\,beam$^{-1}$) is listed in curly brakets. Intensity integrated map contours are in steps of -4$\sigma$, 4$\sigma$, 7$\sigma$, 11$\sigma$ and 13$\sigma$ for C$^{17}$O; and -2$\sigma$, 2$\sigma$, 3$\sigma$, 4$\sigma$, 5$\sigma$, 6$\sigma$ and 7$\sigma$ for C$^{18}$O. PV diagram contours are in steps of -3$\sigma$, 3$\sigma$, 5$\sigma$, 7$\sigma$, 9$\sigma$, 11$\sigma$ and 13$\sigma$ for C$^{17}$O; and -2$\sigma$, 2$\sigma$, 3$\sigma$, 4$\sigma$ and 5$\sigma$ for C$^{18}$O. Keplerian rotation curves (V$\propto$R$^{-0.5}$) for 0.1 and 0.2M$\odot$ and infall curve (V$\propto$R$^{-1}$) are plotted on the PV diagrams. The vertical and horizontal solid lines in the PV diagrams show v$_{lsr}$\,=\,3.8\,km\,s$^{-1}$ and VLA1623A's position, respectively. VLA1623A and VLA1623B are marked by a square and triangle, respectively. (Color figure in the online journal)}
\label{figenv}
\end{figure}

Observed C$^{17}$O and C$^{18}$O emission centered on A present an extended flattened structure perpendicular to the outflow axis (Figure~\ref{figenv}), similar to A's shape in continuum. Both emission show similar velocity gradients characteristic of rotation along the major axis, seen in both velocity gradient maps and Position-Velocity (PV) diagrams (Figure~\ref{figenv}). No velocity gradient is observed along the outflow axis, suggesting contamination from outflows is neglible.

C$^{17}$O's dimensions (FWHM 370AU$\times$240AU) are somewhat larger than A's 1.3mm continuum dimensions (240AU$\times$120AU). Assuming centrally dominated motion, we overlay Keplerian rotation curves (V$\propto$R$^{-0.5}$) for central masses of 0.1 and 0.2M$_{\odot}$ on the C$^{17}$O and C$^{18}$O PV diagrams. For comparison we overlay an infall curve V$\propto$R$^{-1}$. Although the infall curve passes through the peak of C$^{18}$O, both emission are better traced by the Keplerian rotation curves. This suggests that both molecules trace the same rotating disk-like structure, with C$^{18}$O tracing the innermost region and C$^{17}$O the outermost region of the structure due to different resolutions. Though smaller, the characteristics of A's Keplerian disk are similar to \citet{tak12}'s Class I Keplerian disk, hinting at the beginnings of disk evolution. Neither emission appears associated with B and no circumstellar material was detected towards W. 

\section{Discussion}
\subsection{VLA1623A: The prototypical Class 0 source}
From our SED fit of A we obtained T$_{D}$\,=\,36K, higher than previously reported (T$_{D}$\,=\,15--20K, AWB93). This discrepancy may be due to our fitting using only resolved fluxes tracing the inner hotter regions of the core instead of considering the surrounding colder regions, as was done in AWB93. Despite the higher dust temperature, we find A to be a Class 0 object based on lack of detection beyond 24$\mu$m and L$_{submm}$/L$_{bol}$$>$0.5\%, in agreement with previous work. We do not consider the circumstellar-to-protostar-mass ratio as a criterion for evolutionary stage since we cannot account for all of A's envelope mass from our observations.

A presents a flattened elongated envelope structure perpendicular to the outflow direction traced in C$^{17}$O and C$^{18}$O (Figure~\ref{figenv}) as well as in continuum, most noticeable at 1.3mm (Figure~\ref{figcont}). Both C$^{17}$O and C$^{18}$O present signatures of rotation that suggest pure Keplerian rotation (Figure~\ref{figenv}) and provide an estimate for the central object mass of A to be 0.1-0.2M$_{\odot}$. Both the envelope traced by C$^{17}$O and C$^{18}$O and the 1.3mm continuum emission may be tracing a rotating disk about A, a possibility also raised by WKGA11. The presence of this rotating disk about A would be one of the first evidence of Keplerian disks around Class 0 protostars, however further analysis must be carried out to confirm whether this is indeed a Keplerian disk.

\subsection{VLA1623W: A newly identified protostar}
We find W to be a true protostar, based on detection in all IRAC and MIPS1 bands, as well as in the (sub)millimeter and centimeter regimes. We are unable to analyze W's environment due to lack of detection of circumstellar material. Given the lack of circumstellar material emission towards W and the small circumstellar mass M$_{cir}$ derived from the 1.3mm continuum (Table~\ref{tball}), we assume that W's M$_{cir}$ is less than the central object mass M$_{*}$. Additionally, AWB93 showed L$_{sub}$/L$_{bol}$$<$0.5\% to be equivalent to M$_{env}$$<$M$_{*}$. Hence, based on an infrared spectral index $\alpha$$_{IR}$\,=\,1.68$\pm$0.07 (c2d catalogue), L$_{submm}$/L$_{bol}$\,=\,0.18\%, M$_{env}$$<$M$_{\ast}$ and the Class-I-shaped SED, we classify W as a Class I source. \citet{eno09a} mistook W as A+B due to their source identification criteria; however, from the listed source position and $\alpha$$_{IR}$ the source is W, even more so since A lacks infrared emission at $\lambda$$<$24$\mu$m. The fact that W is a protostellar source confirms the suspicions of \citet{den95} and YC97 that the outflow and jet misalignments are caused by a second source.

\subsection{The Nature of VLA1623B}
We find B to be an extremely cold (T$_{D}\sim$5K) object with extremely low luminosity ($\sim$10$^{-5}$L$_{\odot}$). No associated C$^{17}$O and C$^{18}$O emission suggest strong CO depletion, consistent with the derived extremely cold dust temperature (CO sublimation temperature $\sim$20K). We further suspect that B may be driving a pole-on outflow (see section~\ref{secout} and Figure~\ref{figout}). While we cannot determine B's nature unambiguously, some possibilities can be ruled out. 

\citet{cog06} show VLA1623's jet knots have gas temperatures of 1700--4200K and velocities between 60--100\,km\,s$^{-1}$. The strong CO depletion at B indicates that the gas temperature is $<$20K, far below that expected for jet knots. Observations of monopolar continuum jet knots in T Tauri stars \citep{rod12} show obvious shifts in the knot positions in short periods of time. Assuming simple ballistic motion and velocities of 60--100\,km\,s$^{-1}$, B should have shifted from BA97's initially reported position by $\sim$2--3$\arcsec$ in a period of 15 years. However, comparison of BA97's and our 2012 observations show no such shift. VLA high resolution ($\sim$0\farcs5) centimeter continuum observations of jet-driving Class 0 and I objects \citep{rei04} found most detected continuum peaks to be protostellar sources with continuum jets seen as extended emission in the direction of the source's outflow/jet. B is observed as a separate compact component and is coincident with counterparts in the (sub)millimeter regimes, making it more like a YSO and less like a jet knot. Although VLA1623's alignment and distribution might resemble a jet, several MPSs are observed to have their components either aligned or within their companion's outflow (e.g. NGC1333 SVS13: \citealt{che09}; L1448 C: \citealt{hir10}), possibly caused by filament fragmentation. Thus VLA1623's configuration is not highly unlikely. Furthermore, a quick Jeans Mass calculation shows that given B's mass, dust temperature and density, the core is unstable and most likely collapsing. Hence even if it were product of outflow-environment interaction it is on its way to becoming a YSO. Thus, B cannot be a jet/outflow feature as proposed by MOA12.

Other possibilities for the nature of B are a starless core or proto-brown dwarf. While the CO depletion and extremely low temperature and luminosity are characteristics of starless cores \citep{taf02, ai05}, the mass, density and presence of outflow rule out the possibility of B being a starless core. If we assume a star formation efficiency of 30\% the resulting object from B would be a $\sim$0.1M$_{\odot}$ star, well above the brown dwarf limit. With current data and B's physical and chemical characteristics, we cannot determine the eventual outcome but we can conclude that B is an extremely young object between the starless core and Class 0 stages.

\subsection{A triple system \label{secout}}
Comparison of both $^{12}$CO maps and the presence of three sources suggest that VLA1623's outflow is not product of a single source. We propose that the NW lobe is not a cavity but instead two outflows. A drives the northernmost branch while W drives the southermost branch based on its position and the $^{12}$CO (2-1) blueshifted feature. The SE lobe is difficult to disentangle from the observations presented here, requiring comparison of $^{12}$CO and jet observations.

For B, the observed spatially overlapped $^{12}$CO (2-1) emission suggests a compact pole-on outflow. The red- and blueshifted spatially overlapped SO emission and absence of a redshifted lobe in $^{12}$CO (3-2) further support this morphology, since the redshifted lobe of B would be located directly behind the blueshifted lobe which is stronger and more easily detected by the compact configuration observations. Additionally, the $^{12}$CO (3-2) redshifted lobe of B is easily lost among the much stronger redshifted emission of the outflows from A \& W.

Further analysis on VLA1623's kinematics, outflows and whether it is gravitationally bound, as well as determining B's nature, are required to fully disentangle VLA1623. However, from its alignment, physical characteristics, environment and differing evolutionary stages we can conclude that VLA1623 is a triple non-coeval system probably formed through the delayed fragmentation of a filament. These characteristics make VLA1623 a good target for constraining and studying fragmentation in low-mass protostars.


\clearpage

\clearpage

\clearpage

\clearpage

\clearpage

\begin{deluxetable}{ccccc}

\tablecaption{Observations and Physical Parameters of VLA1623}

\tablenum{1}

\tablehead{\colhead{} & \colhead{VLA1623A} & \colhead{VLA1623B} & \colhead{VLA1623W} & \colhead{Ref.}} 

\startdata
RA                   & 16:26:26.397                    & 16:26:26.308    & 16:26:25.628     & 1  \\ 
Dec                  & -24:24:30.71                    & -24:24:30.57    & -24:24:29.60     & 1  \\ 
Separation (\arcsec)\tablenotemark{a} & 0 & 1.18 & 10.51 & 1 \\
\cutinhead{Integrated Flux densities (mJy)\tablenotemark{b}}
3.6$\mu$m        & \nodata                         & \nodata         & 1.9 $\pm$ 0.09    & 2  \\ 
4.5$\mu$m        & \nodata                         & \nodata         & 6.6 $\pm$ 0.2    & 2  \\ 
5.8$\mu$m        & \nodata                         & \nodata         & 0.5 $\pm$ 0.7    & 2  \\ 
8$\mu$m          & \nodata                         & \nodata         & 7.2 $\pm$ 0.2    & 2  \\ 
24$\mu$m         & 40.4 $\pm$ 4.2\tablenotemark{c}  & \nodata         & 272.2 $\pm$ 15.1  & 2  \\ 
870$\mu$m        & 1153 $\pm$ 100\tablenotemark{d}   & 87.2 $\pm$ 12    & 128 $\pm$ 16       & 1  \\ 
1.3mm            & 152 $\pm$ 8                       & 106 $\pm$ 8       & 25.7 $\pm$ 8       & 1  \\ 
2.7mm            & 25.5 $\pm$ 6.3                    & 25.8 $\pm$ 3.5    & 3.1 $\pm$ 1.5\tablenotemark{e}     & 3  \\ 
7mm              & 1.2 $\pm$ 0.2                     & 2.0 $\pm$ 0.2     & \nodata          & 4  \\ 
3.6cm            & 0.1 $\pm$ 0.02                   & 0.4 $\pm$ 0.02   & 0.2 $\pm$ 0.02    & 5  \\ 
4.4cm            &$<$ 0.2                           & $<$ 0.2           & \nodata          & 4  \\ 
6cm              & $<$ 0.4 $\pm$ 0.02                  & $<$ 0.4 $\pm$ 0.02  & 0.2$\pm$0.02    & 4  \\ 
\cutinhead{SED fitting Results}
T$_{D}$ (K)            & 36 $\pm$ 5                      & 5 $\pm$ 1       & 41 $\pm$ 5     & 1  \\ 
$\beta$              & 1.2 $\pm$ 0.6                   & 1.7 $\pm$ 0.5   & 1.6 $\pm$ 0.4    & 1  \\
Reduced $\chi$$^{2}$        & 21.7                            & 5.8             & 0.7              & 1  \\
 \cutinhead{Physical Parameters}
L$_{bol}$ (L$_{\odot}$) & 0.03 $\pm$ 0.003 & (8.9 $\pm$ 0.5) $\times$ 10$^{-5}$ & 0.03 $\pm$ 0.002 & 1 \\
L$_{submm}$ (L$_{\odot}$) & (3.9 $\pm$ 0.3)$\times$ 10$^{-4}$ & \nodata & (4.8 $\pm$ 0.5) $\times$ 10$^{-5}$ & 1 \\
L$_{submm}$/L$_{bol}$ ($\%$) & 1.2 $\pm$ 0.1 & \nodata & 0.2 $\pm$ 0.02 & 1 \\
L$_{bol,SED}$ (L$_{\odot}$) & 1.1 $\pm$ 0.2 & (2.9 $\pm$ 0.1) $\times$ 10$^{-4}$ & 1.4 $\pm$ 0.2 & 1 \\
L$_{submm,SED}$ (L$_{\odot}$) & (1 $\pm$ 0.6)$\times$ 10$^{-2}$ & \nodata & (5 $\pm$ 2) $\times$ 10$^{-3}$ & 1 \\
L$_{submm,SED}$/L$_{bol,SED}$ ($\%$) & 0.9 $\pm$ 0.5 & \nodata & 0.4 $\pm$ 0.2 & 1 \\
M$_{cir}$ (M$_{\odot}$)\tablenotemark{f} & 0.02 $\pm$ 0.004 & 0.3 $\pm$ 0.2 & 0.003 $\pm$ 0.001 & 1 \\
N$_{H_{2}}$ (cm$^{-2}$)\tablenotemark{f} & (4.4 $\pm$ 0.8) $\times$ 10$^{23}$ & (9.9 $\pm$ 5.0) $\times$ 10$^{24}$ & (3.6 $\pm$ 1.2) $\times$ 10$^{23}$ & 1 \\
n$_{H_{2}}$ (cm$^{-3}$)\tablenotemark{f} & (2.2 $\pm$ 0.4) $\times$ 10$^{8}$ & (5.5 $\pm$ 2.8) $\times$ 10$^{9}$ & (4.1 $\pm$ 1.4) $\times$ 10$^{8}$ & 1 \\
\enddata

\tablenotetext{a}{Separation with respect to VLA1623A}
\tablenotetext{b}{Flux uncertainties are statistical}
\tablenotetext{c}{Unresolved VLA1623A \& B flux. However, since VLA1623B peaks at around 1000$\mu$m, its contribution to the 24$\mu$m flux is neglible.}
\tablenotetext{d}{Flux calculated by subtracting VLA1623B's 870$\mu$m extended integrated flux from the unresolved VLA1623A \& B 870$\mu$m 870$\mu$m compact flux.}
\tablenotetext{e}{We obtained reduced BIMA data from \citet{lmw00} and calculated the flux ourselves.}
\tablenotetext{f}{Calculated from 1.3mm continuum emission}

\tablerefs{(1)This work; (2)c2d Catalogue; (3)LMW00; (4)WKGA11; (5)BA97}
\label{tball}
\end{deluxetable}

\end{document}